# Probing the Higgs mechanism via $\gamma\gamma \to \mathrm{W}^+\mathrm{W}^-$

A. Denner, R. Schuster†

*Institut für Theoretische Physik, Universität Würzburg*
*Am Hubland, D-97074 Würzburg, Germany*

S. Dittmaier‡

*Theoretische Physik, Universität Bielefeld*
*Universitätsstraße, D-33501 Bielefeld, Germany*

**Abstract:**

We investigate the sensitivity of the reaction $\gamma\gamma \to \mathrm{W}^+\mathrm{W}^-$ to the Higgs sector based on the complete one-loop corrections in the minimal Standard Model and the gauged non-linear $\sigma$-model. While this sensitivity is very strong for the suppressed cross-section of equally polarized photons and longitudinal W bosons, it is only marginal for the dominant mode of transverse polarizations. The corrections within the $\sigma$-model turn out to be UV-finite in accordance with the absence of $\log M_{\mathrm{H}}$ terms in the Standard Model with a heavy Higgs boson.

†Supported by the Deutsche Forschungsgemeinschaft.

‡Supported by the Bundesministerium für Forschung und Technologie, Bonn, Germany.

## 1 Introduction

All present experimental results on electroweak physics confirm the conception that electromagnetic and weak interactions are unified in a $SU(2) \times U(1)$ gauge theory. However, the underlying field theory cannot be of pure Yang-Mills type since the weak gauge bosons, the $W^{\pm}$ and Z boson, are empirically known to be massive. In the electroweak Standard Model (SM) this problem is solved by the well-known Higgs mechanism [1], i.e. by breaking the gauge symmetry spontaneously via a non-vanishing vacuum expectation value of an additional complex scalar SU(2) doublet. Whereas three of these four scalar fields are absorbed by the longitudinal degrees of freedom of the massive gauge bosons, a physical scalar field survives, the so-called Higgs boson. Of course, the Higgs mechanism cannot be conclusively confirmed before this particle is empirically detected. On the other hand, the Higgs-boson mass $M_{\mathrm{H}}$, which is a free parameter of the theory, enters all theoretical predictions within the SM at least via higher orders. Since the Higgs-mass dependence of low-energy observables turns out to be very mild, more precisely at most logarithmic at the one-loop level, only crude bounds on $M_{\mathrm{H}}$ can be obtained from radiative corrections (RCs) to current precision measurements. Experimentally, the Higgs mass is only constrained by the lower bound $M_{\mathrm{H}} \gtrsim 60\,\mathrm{GeV}$ from LEP [2] but can well be in the TeV range.

The Higgs boson can be removed from the physical particle spectrum in two different ways. On the one hand, amplitudes can be calculated within the SM for finite $M_{\mathrm{H}}$, and subsequently asymptotically expanded for $M_{\mathrm{H}} \to \infty$. Alternatively, the physical Higgs field can be eliminated by constraining the square of the Higgs-doublet field to be constant and equal to its (non-vanishing) vacuum expectation value. Then no physical Higgs particle exists from the beginning, but one is forced to introduce a non-linear representation of the Higgs sector leading to a non-renormalizable gauged non-linear $\sigma$-model (GNLSM). The relation between the heavy-Higgs limit and the GNLSM has been investigated for a SU(2) gauge theory and the $SU(2) \times U(1)$ SM in Refs. [3] and [4,5], respectively. As expected, $M_{\mathrm{H}}$ acts as an effective UV cut-off. The corresponding (logarithmic) one-loop divergences in the GNLSM can be identified with the $\log M_{\mathrm{H}}$ terms in the SM only up to finite constants, which have been calculated in Ref. [5]. Although the GNLSM is manifestly non-renormalizable, and its observables in general violate unitarity in the high-energy limit, an investigation of the GNLSM seems reasonable since it is equivalent to the SM in the unitary gauge with the physical Higgs field omitted. Consequently, by comparing theoretical predictions within the GNLSM and the SM for varying $M_{\mathrm{H}}$ one may get insight into the influence of the mechanism of spontaneous symmetry breaking on specific observables. The discussion of these aspects for the cross-section of $\gamma\gamma \to W^+W^-$ represents the main issue of this paper.

The process $\gamma\gamma \to W^+W^-$ will be one of the most important reactions at future $\gamma\gamma$ colliders. In particular, the measurement of the corresponding cross-section yields direct information on possible anomalous $\gamma$WW and $\gamma\gamma$WW couplings [6] widely independent of the couplings between Z and $W^{\pm}$ bosons. Moreover, a Higgs boson with a mass of several hundred GeV can be studied via the resonance contribution $\gamma\gamma \to H^* \to W^+W^-$, which is present owing to the $\gamma\gamma$H coupling induced at one-loop order. Since the structure of this Higgs resonance has already been discussed in the literature [7,8], here we mainly concentrate on the case when the centre-of-mass energy is far below the Higgs mass $M_{\rm H}$.

We have calculated the full one-loop RCs to $\gamma\gamma \to {\rm W}^+{\rm W}^-$ including soft-photon bremsstrahlung both in the SM and GNLSM. A complete discussion of the SM RCs will be published elsewhere [9]; here we focus on the $M_{\rm H}$ dependence of the SM corrections and their difference to the ones within the GNLSM. Despite of the non-renormalizability of the GNLSM, the corresponding one-loop RCs to $\gamma\gamma \to {\rm W}^+{\rm W}^-$ turn out to be ultraviolet finite. This fact is related to the absence of $\log M_{\rm H}$ terms in the SM corrections. The limit $M_{\rm H} \to \infty$ indeed exists for the SM one-loop corrections, but for longitudinal polarized W bosons these one-loop corrected cross-sections violate unitarity for energies in the TeV range, as it is also the case in the GNLSM.

The paper is organized as follows: In Section 2 we discuss the $M_{\rm H}$ dependence of the SM RCs and their difference to the ones within the GNLSM. The unitarity-violating effects for longitudinal W bosons are investigated in Section 3. Numerical results are presented in Section 4. Section 5 contains our conclusions.

## 2 Heavy-Higgs Standard Model versus gauged non-linear $\sigma$-model

The GNLSM is related (see e.g. Ref. [10]) to the SM in the unitary gauge without Higgs field by a Stueckelberg transformation [11]. Comparing the Lagrangians, one finds that the Feynman rules involving at most one unphysical scalar field are identical in the GNLSM and the SM with linearly realized Higgs sector. Vertices with at least two scalar fields are in general different. In particular, the ${\rm WW}\varphi\varphi$ and ${\rm WW}\chi\chi$ couplings vanish in the GNLSM. By $\varphi$ and $\chi$ we denote the charged and neutral unphysical scalar fields, respectively. For the reaction $\gamma\gamma \to {\rm W}^+{\rm W}^-$ at one loop one simply has to omit all graphs that contain internal Higgs fields, or ${\rm WW}\varphi\varphi$ or ${\rm WW}\chi\chi$ couplings in order to obtain the GNLSM results from the SM ones.

Obviously, the tree-level amplitudes agree in both models yielding

$$\begin{aligned}
\mathcal{M}_{\rm Born} = 4\pi\alpha\Biggl\{ \frac{2}{M_{\rm W}^2 - t}\Bigl[ & 2(\varepsilon_1\cdot\varepsilon_2)(k_1\cdot\varepsilon_+^*)(k_2\cdot\varepsilon_-^*) + 2(\varepsilon_+^*\cdot\varepsilon_-^*)(k_+\cdot\varepsilon_1)(k_-\cdot\varepsilon_2) \\
& - 2(\varepsilon_1\cdot\varepsilon_+^*)(k_1\cdot\varepsilon_-^*)(k_-\cdot\varepsilon_2) - 2(\varepsilon_2\cdot\varepsilon_-^*)(k_2\cdot\varepsilon_+^*)(k_+\cdot\varepsilon_1) \\
& + 2(\varepsilon_1\cdot\varepsilon_-^*)(k_1\cdot\varepsilon_+^*)(k_-\cdot\varepsilon_2) + 2(\varepsilon_2\cdot\varepsilon_+^*)(k_2\cdot\varepsilon_-^*)(k_+\cdot\varepsilon_1) \\
& + s(\varepsilon_1\cdot\varepsilon_+^*)(\varepsilon_2\cdot\varepsilon_-^*)\Bigr] - (\varepsilon_1\cdot\varepsilon_2)(\varepsilon_+^*\cdot\varepsilon_-^*)\Biggr\} \\
+ (\text{`1'} \leftrightarrow \text{`2'}, t \to u). & \qquad (1)
\end{aligned}$$

The momenta $k$ and polarization vectors $\varepsilon$ of the incoming photons are labelled by '1','2', the ones of the outgoing ${\rm W}^\pm$ bosons by '$\pm$', respectively; they are explicitly defined in the centre-of-mass (CM) system in Ref. [9]. The Mandelstam variables are given by

$$s = (k_1+k_2)^2 = 4E^2, \quad t = (k_1-k_+)^2 = M_{\rm W}^2 - \frac{s}{2}(1-\beta\cos\theta), \quad u = 2M_{\rm W}^2 - s - t, \qquad (2)$$

with $\beta = \sqrt{1 - M_{\mathrm{W}}^2/E^2}$ denoting the velocity of the W bosons, and $\theta$ representing the scattering angle between photon '1' and $\mathrm{W}^+$.

For longitudinal W bosons the lowest-order matrix elements read explicitly

$$\mathcal{M}_{\mathrm{Born}}(\lambda_1 = \lambda_2, \lambda_\pm = 0) = 4\pi\alpha \frac{2sM_{\mathrm{W}}^2}{(M_{\mathrm{W}}^2 - t)(M_{\mathrm{W}}^2 - u)} \underset{|q^2| \gg M_{\mathrm{W}}^2}{\sim} 4\pi\alpha \frac{2sM_{\mathrm{W}}^2}{ut},$$
$$\mathcal{M}_{\mathrm{Born}}(\lambda_1 = -\lambda_2, \lambda_\pm = 0) = -4\pi\alpha \frac{2(s + 4M_{\mathrm{W}}^2)(ut - M_{\mathrm{W}}^4)}{s\beta^2(M_{\mathrm{W}}^2 - t)(M_{\mathrm{W}}^2 - u)} \underset{|q^2| \gg M_{\mathrm{W}}^2}{\sim} -8\pi\alpha. \qquad (3)$$

Here and in the following we denote the Mandelstam variables $s, t, u$ generically by $q^2$. Note that in the high-energy limit $|q^2| \gg M_{\mathrm{W}}^2$ the amplitude for equal photon helicities vanishes and that the other one contains no $t$- and $u$-channel pole in the leading term.

The calculation of the one-loop amplitude for $\gamma\gamma \to \mathrm{W}^+\mathrm{W}^-$ is simplified considerably by use of a non-linear gauge-fixing condition for the W-boson field, suggested in Ref. [12], rendering the $\gamma\varphi\mathrm{W}$ coupling zero [9]. Using this gauge-fixing condition in the GNLSM as well, we have evaluated the difference of the one-loop matrix elements for $\gamma\gamma \to \mathrm{W}^+\mathrm{W}^-$ in the SM and the GNLSM

$$\delta\mathcal{M}_{\mathrm{H}} = \delta\mathcal{M}_{\mathrm{SM}} - \delta\mathcal{M}_{\mathrm{GNLSM}}, \qquad (4)$$

where $\delta\mathcal{M}$ always denotes one-loop contributions to the amplitude. In the limit of very large Higgs mass, $M_{\mathrm{H}}^2 \gg |q^2|, M_{\mathrm{W}}^2$ our result simplifies to

$$\begin{aligned} \delta\mathcal{M}_{\mathrm{H}}\Big|_{M_{\mathrm{H}}\to\infty} = \frac{\alpha^2}{3s_{\mathrm{W}}^2(M_{\mathrm{W}}^2 - t)} \Big\{ & s(\varepsilon_1 \cdot \varepsilon_+^*)(\varepsilon_2 \cdot \varepsilon_-^*) + 2(\varepsilon_1 \cdot \varepsilon_2)(k_1 \cdot \varepsilon_+^*)(k_2 \cdot \varepsilon_-^*) \\ & - (\varepsilon_1 \cdot \varepsilon_+^*)(k_1 \cdot \varepsilon_-^*)(k_- \cdot \varepsilon_2) + (\varepsilon_1 \cdot \varepsilon_-^*)(k_1 \cdot \varepsilon_+^*)(k_- \cdot \varepsilon_2) \\ & - (\varepsilon_2 \cdot \varepsilon_-^*)(k_2 \cdot \varepsilon_+^*)(k_+ \cdot \varepsilon_1) + (\varepsilon_2 \cdot \varepsilon_+^*)(k_2 \cdot \varepsilon_-^*)(k_+ \cdot \varepsilon_1) \Big\} \\ & + (\text{`1'} \leftrightarrow \text{`2'}, t \to u), \end{aligned} \qquad (5)$$

whereas the exact analytical form of $\delta\mathcal{M}_{\mathrm{H}}$ for arbitrary Higgs mass is not very illuminating. In this context, we mention that we have derived (5) also using the effective Lagrangian for the difference of the SM limit $M_{\mathrm{H}} \to \infty$ and the GNLSM given in Refs. [4,5].[1]

The logarithmic one-loop UV divergences occuring in the non-renormalizable GNLSM are directly related to the $\log M_{\mathrm{H}}$ terms in the SM with a heavy Higgs boson, i.e. $M_{\mathrm{H}}$ can be regarded as an effective UV cut-off in this limit. This fact has already been pointed out in Ref. [4] and shown by explicit calculation in Ref. [5]. Thus, the absence of $\log M_{\mathrm{H}}$ terms in $\delta\mathcal{M}_{\mathrm{SM}}$ and the UV finiteness of $\delta\mathcal{M}_{\mathrm{GNLSM}}$ have the same root, however, the difference $\delta\mathcal{M}_{\mathrm{H}}$ is non-vanishing even for $M_{\mathrm{H}} \to \infty$. Of course, all results derived in non-renormalizable models are not free from ambiguities or assumptions that fix these ambiguities so that such results have to be interpreted carefully. But the finiteness of $\delta\mathcal{M}_{\mathrm{GNLSM}}$ shows that the prediction for $\gamma\gamma \to \mathrm{W}^+\mathrm{W}^-$ within the GNLSM is independent of any cut-off $\Lambda$ whatever regularization procedure may be used. Note that such a $\Lambda$ will play a role as "scale of new physics" if the GNLSM is embedded into a more complete field theory like the SM (where $\Lambda \sim M_{\mathrm{H}}$) or even beyond. Moreover, the difference $\delta\mathcal{M}_{\mathrm{H}}$ indicates to which extent the SM prediction might be modified by effects of new physics concerning the Higgs sector.

[1]More precisely, some missing counterterms involving $\alpha_{11}$ had to be supplemented in the Feynman rules of Ref. [4], and the finite parts of the contributing $\alpha_i$ could be taken from Ref. [5].

## 3 Production of longitudinal W bosons in the high-energy limit

Observables involving longitudinally polarized massive gauge bosons are most sensitive to deviations from the Yang-Mills interactions and the mechanism of spontaneous symmetry breaking of the underlying gauge theory for energies far above the scale of the gauge-boson masses. This is due to the well-known "gauge cancellations" which guarantee that the enhancement factor $E/M$ of the longitudinal polarization vector (of a vector boson with energy $E$ and mass $M$) is cancelled between the individual contributions to the S-matrix elements. The equivalence theorem (ET) [13] states that in the SM the leading contribution to amplitudes involving external longitudinally polarized gauge bosons can be simply obtained by the replacement of this vector field by the corresponding unphysical scalar field, if all energy scales $q_i^2$ are far above all masses $m_i$, $|q_i^2| \gg m_i^2$. Moreover, the ET can be generalized to the heavy-Higgs SM, $|q_i^2|, M_{\mathrm{H}}^2 \gg m_i^2$, and the GNLSM [14,15].

Applying the ET to $\gamma\gamma \to \mathrm{W}_{\mathrm{L}}^+\mathrm{W}_{\mathrm{L}}^-$ within the heavy-Higgs SM and the GNLSM, all one-loop RCs of the order $M_{\mathrm{H}}^2/M_{\mathrm{W}}^2$, $q^2/M_{\mathrm{W}}^2$ ($q^2 = s,t,u$) can be obtained from Feynman diagrams of $\gamma\gamma \to \varphi^+\varphi^-$ involving only scalar inner particles, as can be deduced by power counting [15]. In the heavy-Higgs SM these diagrams are shown and calculated in Ref. [7] for equal photon helicities. The result for $\gamma\gamma \to \mathrm{W}_{\mathrm{L}}^+\mathrm{W}_{\mathrm{L}}^-$ with general photon helicities is given by

$$
\begin{aligned}
\delta\mathcal{M}_{\mathrm{SM}}(\lambda_1=\lambda_2,\lambda_\pm=0) \sim & -\frac{\alpha^2 M_{\mathrm{H}}^2}{2s_{\mathrm{W}}^2 M_{\mathrm{W}}^2}\left\{\frac{M_{\mathrm{H}}^2}{s-M_{\mathrm{H}}^2+iM_{\mathrm{H}}\Gamma_{\mathrm{H}}}+2\right\} \\
& +\frac{\alpha^2 M_{\mathrm{H}}^2}{2s_{\mathrm{W}}^2 M_{\mathrm{W}}^2}\left\{\frac{M_{\mathrm{H}}^2}{t}\left(\frac{M_{\mathrm{H}}^2}{t}-1\right)\log\left(1-\frac{t}{M_{\mathrm{H}}^2}\right)+\frac{M_{\mathrm{H}}^2}{t} \quad + \quad (t\leftrightarrow u)\right\}, \qquad (6)
\end{aligned}
$$

$$
\begin{aligned}
\delta\mathcal{M}_{\mathrm{SM}}(\lambda_1=-\lambda_2,\lambda_\pm=0) \sim & -\frac{\alpha^2 M_{\mathrm{H}}^2}{s_{\mathrm{W}}^2 M_{\mathrm{W}}^2} \\
& \times\left\{\frac{M_{\mathrm{H}}^2(M_{\mathrm{H}}^2-t)}{2ut}\left[\log\left(\frac{-s-i\epsilon}{M_{\mathrm{H}}^2}\right)\log\left(1-\frac{t}{M_{\mathrm{H}}^2}\right)+\mathrm{Li}_2\left(\frac{t}{M_{\mathrm{H}}^2}\right)-\frac{u}{t}\log\left(1-\frac{t}{M_{\mathrm{H}}^2}\right)\right]\right. \\
& \left.-\frac{M_{\mathrm{H}}^2}{2t}+\frac{1}{8}+\frac{M_{\mathrm{H}}^2(u-s-2M_{\mathrm{H}}^2)}{4st}\left[\mathrm{Li}_2\left(1+\frac{s+i\epsilon}{M_{\mathrm{H}}^2}\right)-\frac{\pi^2}{6}\right] \quad + \quad (t\leftrightarrow u)\right\}, \qquad (7)
\end{aligned}
$$

$$
\text{for} \quad s,-t,-u,M_{\mathrm{H}}^2 \gg M_{\mathrm{W}}^2,
$$

where (6) is in agreement with Ref. [7].[2] The graphs which are relevant in the GNLSM are shown in Fig. 1. Note that the corresponding $\varphi\varphi\varphi\varphi$ coupling in the GNLSM implicitly

[2]The difference in the global sign is due to deviating phase conventions for the polarization vectors.

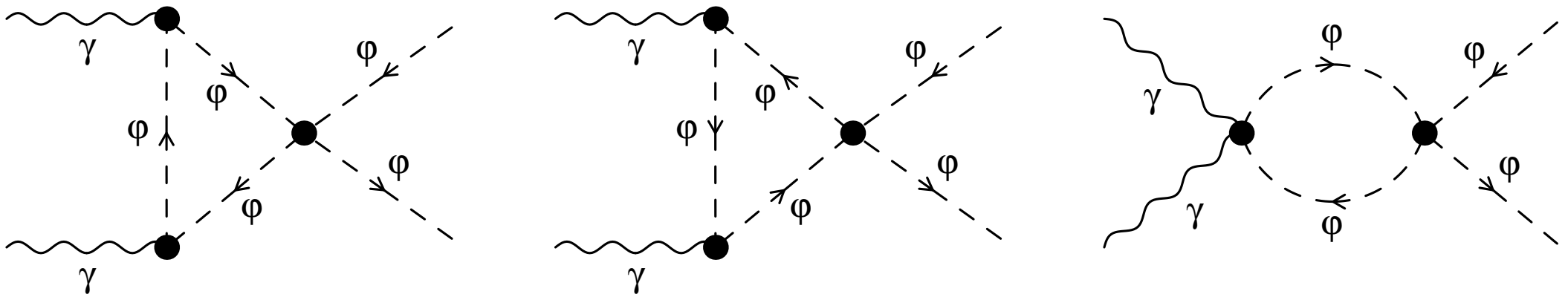

Figure 1: Feynman diagrams for $\gamma\gamma \to \varphi^+\varphi^-$ in the GNLSM relevant for the leading high-energy behaviour.

contains enhancement factors of the type $q^2/M_\mathrm{W}^2$. The final result reads

$$\delta\mathcal{M}_\mathrm{GNLSM}(\lambda_\pm = 0) \sim \frac{\alpha^2}{4s_\mathrm{W}^2}\frac{s}{M_\mathrm{W}^2}\delta_{\lambda_1\lambda_2} \qquad \text{for} \quad |q^2| \gg M_\mathrm{W}^2. \tag{8}$$

Using (6) and (7) we can compare the $q^2/M_\mathrm{W}^2$ terms of the GNLSM with the corresponding SM limit $M_\mathrm{H} \to \infty$, given by

$$\delta\mathcal{M}_\mathrm{SM}(\lambda_\pm = 0)\Big|_{M_\mathrm{H}\to\infty} \sim \frac{5\alpha^2}{12s_\mathrm{W}^2}\frac{s}{M_\mathrm{W}^2}\delta_{\lambda_1\lambda_2} \qquad \text{for} \quad |q^2| \gg M_\mathrm{W}^2. \tag{9}$$

Consequently, even the unitarity-violating $s/M_\mathrm{W}^2$ terms are different in the SM with $M_\mathrm{H} \to \infty$ and the GNLSM. Of course, these terms are absent in the high-energy limit of the SM if $M_\mathrm{H}$ is kept finite, i.e. $|q^2| \gg M_\mathrm{H}^2 \gg M_\mathrm{W}^2$. In this case (6) and (7) reduce to

$$\delta\mathcal{M}_\mathrm{SM}(\lambda_1 = \lambda_2, \lambda_\pm = 0) \sim -\frac{\alpha^2 M_\mathrm{H}^2}{s_\mathrm{W}^2 M_\mathrm{W}^2}, \qquad \text{for} \quad |q^2| \gg M_\mathrm{H}^2 \gg M_\mathrm{W}^2, \tag{10}$$

$$\delta\mathcal{M}_\mathrm{SM}(\lambda_1 = -\lambda_2, \lambda_\pm = 0) \sim -\frac{\alpha^2 M_\mathrm{H}^2}{8s_\mathrm{W}^2 M_\mathrm{W}^2}, \qquad \text{for} \quad |q^2| \gg M_\mathrm{H}^2 \gg M_\mathrm{W}^2. \tag{11}$$

The way how these various leading corrections influence the complete one-loop RCs can be seen in the numerical discussion of the next section.

## 4 Numerical results

For the numerical evaluations we use the parameters of Ref. [2]. In particular, the W-boson mass is kept fixed to $M_\mathrm{W} = 80.22\,\mathrm{GeV}$. All integrated cross-sections are obtained from the angular range $10° < \theta < 170°$. The polarizations of the external particles are indicated by four labels, the first two corresponding to the photons and the last two to the W bosons. The label U stands for unpolarized, + for right-handed, − for left-handed, T for transverse and L for longitudinal. Since we are interested only in the $M_\mathrm{H}$ dependence of the SM RCs and their difference to the ones within the GNLSM we omit all $\log(\Delta E/E)$ terms, which represent the cut-off-dependent corrections originating from soft-bremsstrahlung photons of energy $E_\gamma < \Delta E$. As already mentioned, a more complete discussion of the SM RCs to $\gamma\gamma \to \mathrm{W}^+\mathrm{W}^-$ will be published elsewhere [9].

In order to set the scale, we first show in Fig. 2 the lowest-order integrated cross-sections for various polarizations. At high energies, the unpolarized cross-section $\sigma_\mathrm{UUUU}^\mathrm{Born}$ is dominated by transverse W bosons and all polarized cross-sections involving two transverse W bosons are of the same order. The cross-sections involving longitudinal W bosons are smaller owing to the suppression of the $t$- and $u$-channel pole [see (3)]. While the cross-sections for opposite photon polarizations and mixed transverse and longitudinal W-boson polarizations $\sigma^{\rm Born}_{\pm\mp{\rm TL}}$ and $\sigma^{\rm Born}_{\pm\mp{\rm LT}}$ are suppressed by an additional factor $1/s$, the corresponding ones for equal photon helicities vanish at lowest order. Finally, the cross-section for equal photon helicities and purely longitudinal W bosons $\sigma^{\rm Born}_{\pm\pm{\rm LL}}$, the most interesting one for the study of the Higgs sector, behaves like $1/s^3$ at high energies and is suppressed with respect to the unpolarized cross-section by more than four orders of magnitude already at $E_{\rm CMS} = 1\,{\rm TeV}$.

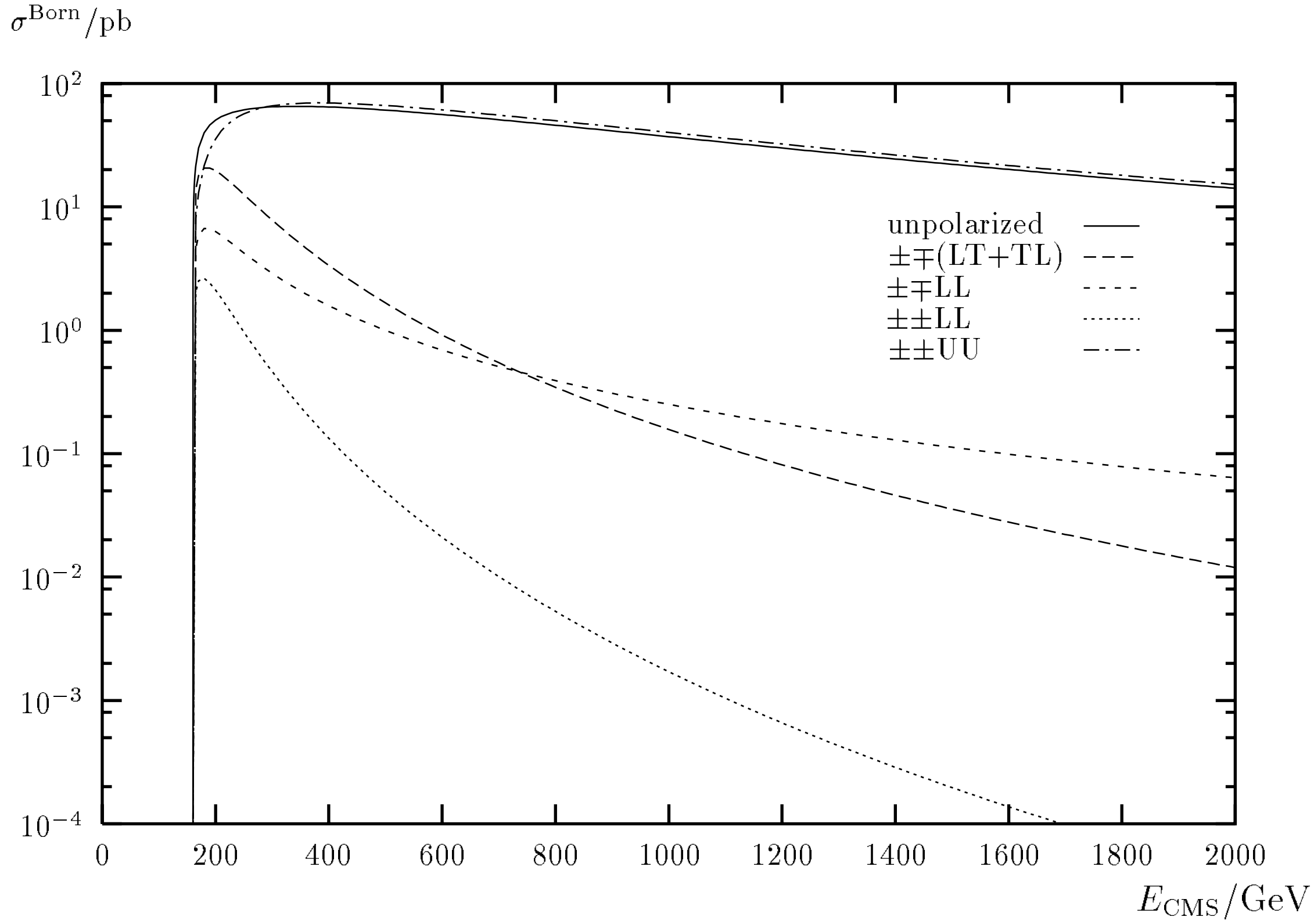

Figure 2: Lowest-order integrated cross-section for various polarizations.

Owing to the strong suppression of the lowest-order cross-section and the presence of unitarity-violating effects in the $\mathcal{O}(\alpha)$ corrections, the cross-section $\sigma_{\pm\pm{\rm LL}}$ is dominated by the $\mathcal{O}(\alpha)$ corrections at high energies. Consequently, we have calculated the corrected cross-section for this polarization by squaring the complete matrix element so that the large relative corrections of order $\mathcal{O}(\alpha^2 s^2/M_{\rm W}^4)$ are treated properly. Squaring also the non-leading one-loop RCs changes the result only at the order of the neglected two-loop corrections. In order to get an IR-finite result we have also squared the real soft-bremsstrahlung correction, which is proportional to the lowest-order cross-section and thus very small. For all other polarized cross-sections we include only the strict $\mathcal{O}(\alpha)$ corrections, i.e. the interference of the corrections with the lowest-order amplitude but not the square of the $\mathcal{O}(\alpha)$ corrections. The unpolarized cross-sections are obtained by summing the polarized ones calculated as described above.

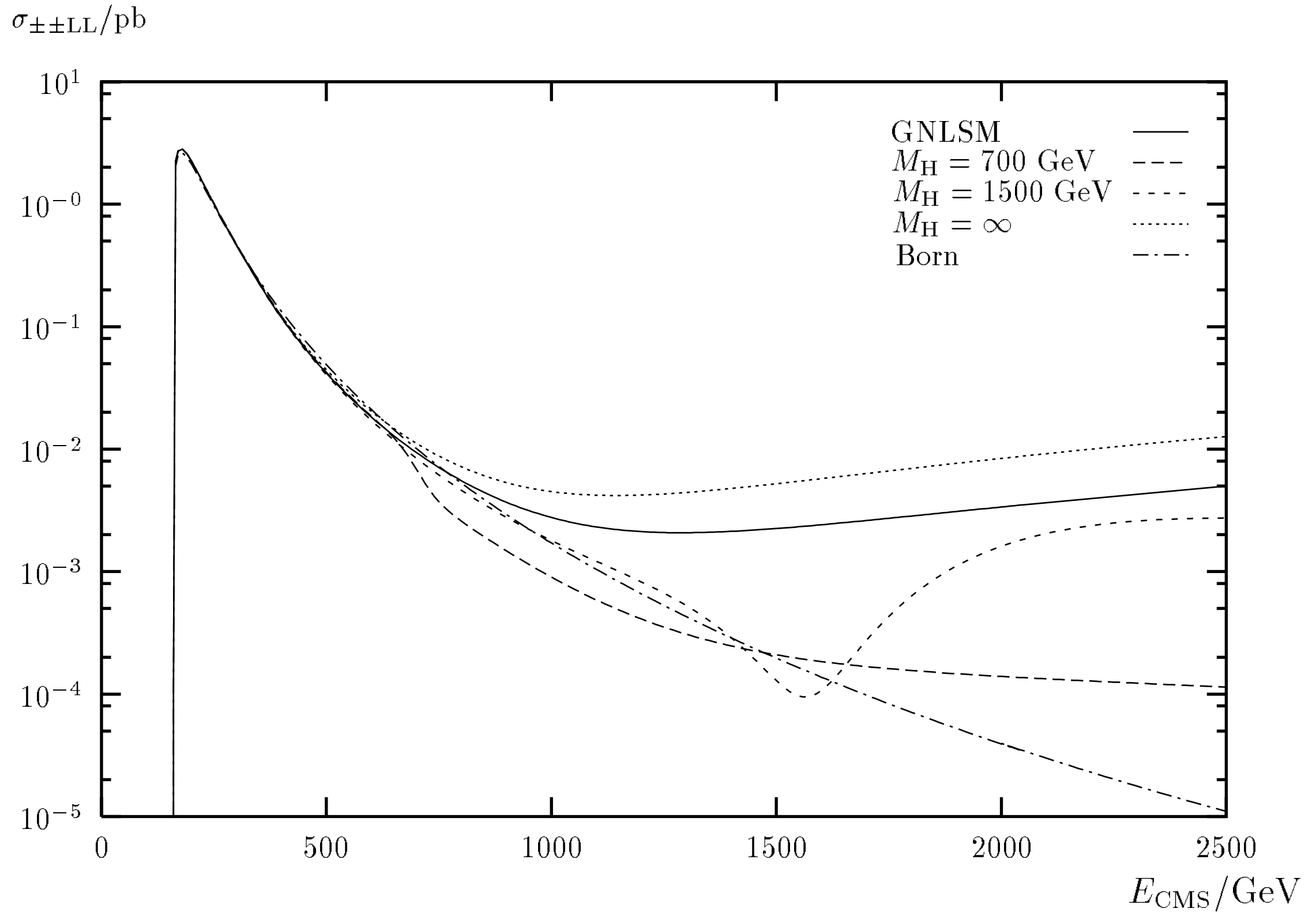

Figure 3: Integrated cross-section for equal photon helicities and purely longitudinal W bosons.

The strong enhancement of $\sigma_{\pm\pm\mathrm{LL}}$ arising from the higher-order corrections is demonstrated in Fig. 3. For large Higgs masses it amounts to more than two orders of magnitude at 2 TeV. For high energies the cross-section depends very strongly on the Higgs-boson mass. While for $M_{\mathrm{H}} \gg s$ the behaviour of the cross-section is governed by (9), for $M_{\mathrm{H}} \ll s$ it is given by (10). The two regions are separated by the Higgs resonance. The cross-section for $M_{\mathrm{H}} = \infty$ grows with $s$ and eventually violates unitarity. It deviates from the one of the GNLSM by a factor of roughly 9/25 at high energies in accordance with (8) and (9). Figure 3 qualitatively agrees with the one shown in Ref. [7] where only the enhanced terms of order $\mathcal{O}(\alpha M_{\mathrm{H}}^2/M_{\mathrm{W}}^2)$ and $\mathcal{O}(\alpha s/M_{\mathrm{W}}^2)$ for $\gamma\gamma \to \mathrm{W}_{\mathrm{L}}^+\mathrm{W}_{\mathrm{L}}^-$ were calculated. The strong sensitivity of $\sigma_{\pm\pm\mathrm{LL}}$ to the Higgs sector will probably be very hard to exploit in the presence of the enormous background of transverse W-boson production.

For the other polarizations, the relative corrections $\delta = \sigma/\sigma_{\mathrm{Born}} - 1$ to the integrated cross-section are illustrated in Figs. 4 and 5. Note that in all those cross-sections no unitarity-violating terms appear for $M_{\mathrm{H}} = \infty$ in the SM or in the GNLSM and that for finite but not very small $M_{\mathrm{H}}$ no Higgs resonance is visible. More precisely, the Higgs resonance is only present for equally polarized photons but suppressed for transverse W bosons. While for the polarizations involving longitudinal W bosons a dependence on the Higgs mass of nearly 10% shows up (Fig. 4), such a dependence is not visible for purely transverse W bosons and all cross-sections including these polarizations (Fig. 5).

The background of transverse W bosons can be reduced by more stringent angular cuts. This can most easily be seen by considering the differential cross-sections plotted in Figs. 6 and 7 for $E_{\mathrm{CMS}} = 2\,\mathrm{TeV}$. While the angular distribution for longitudinal

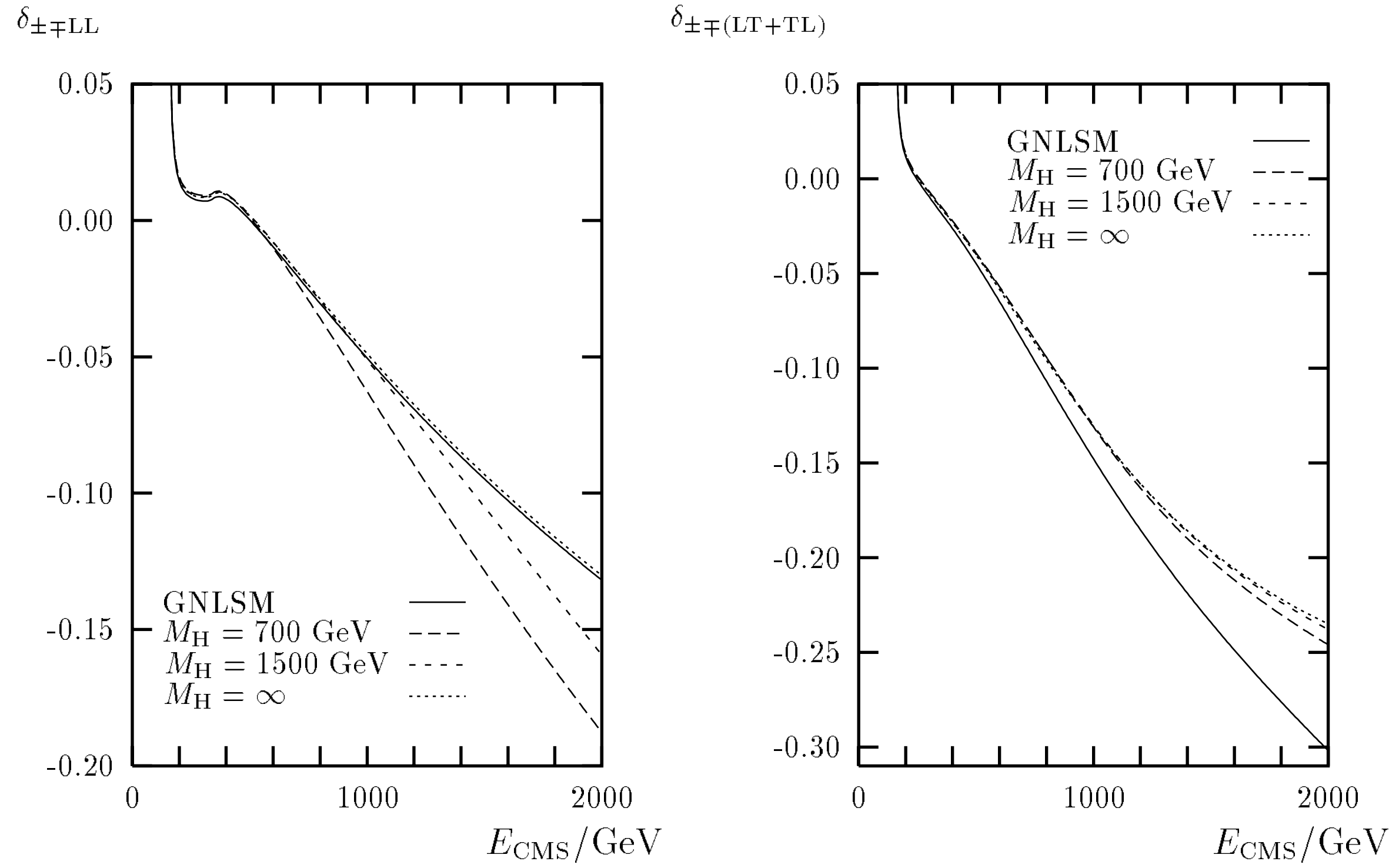

Figure 4: Corrections to the integrated cross-section for unequal photon helicities and purely longitudinal W bosons (left) or mixed transverse and longitudinal W bosons (right).

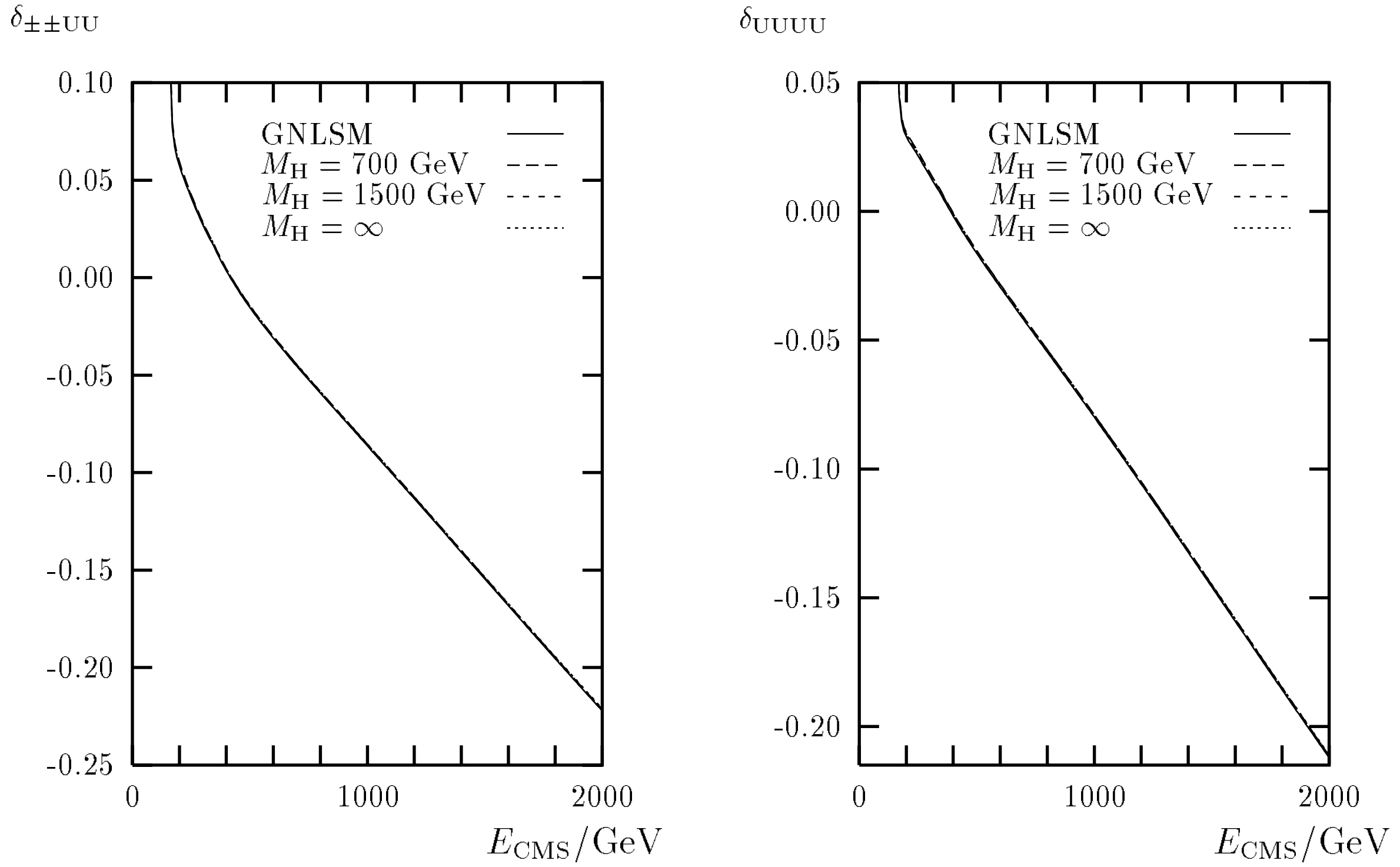

Figure 5: Corrections to the integrated cross-section for unpolarized W bosons and equally polarized photons (left) or unpolarized photons (right).

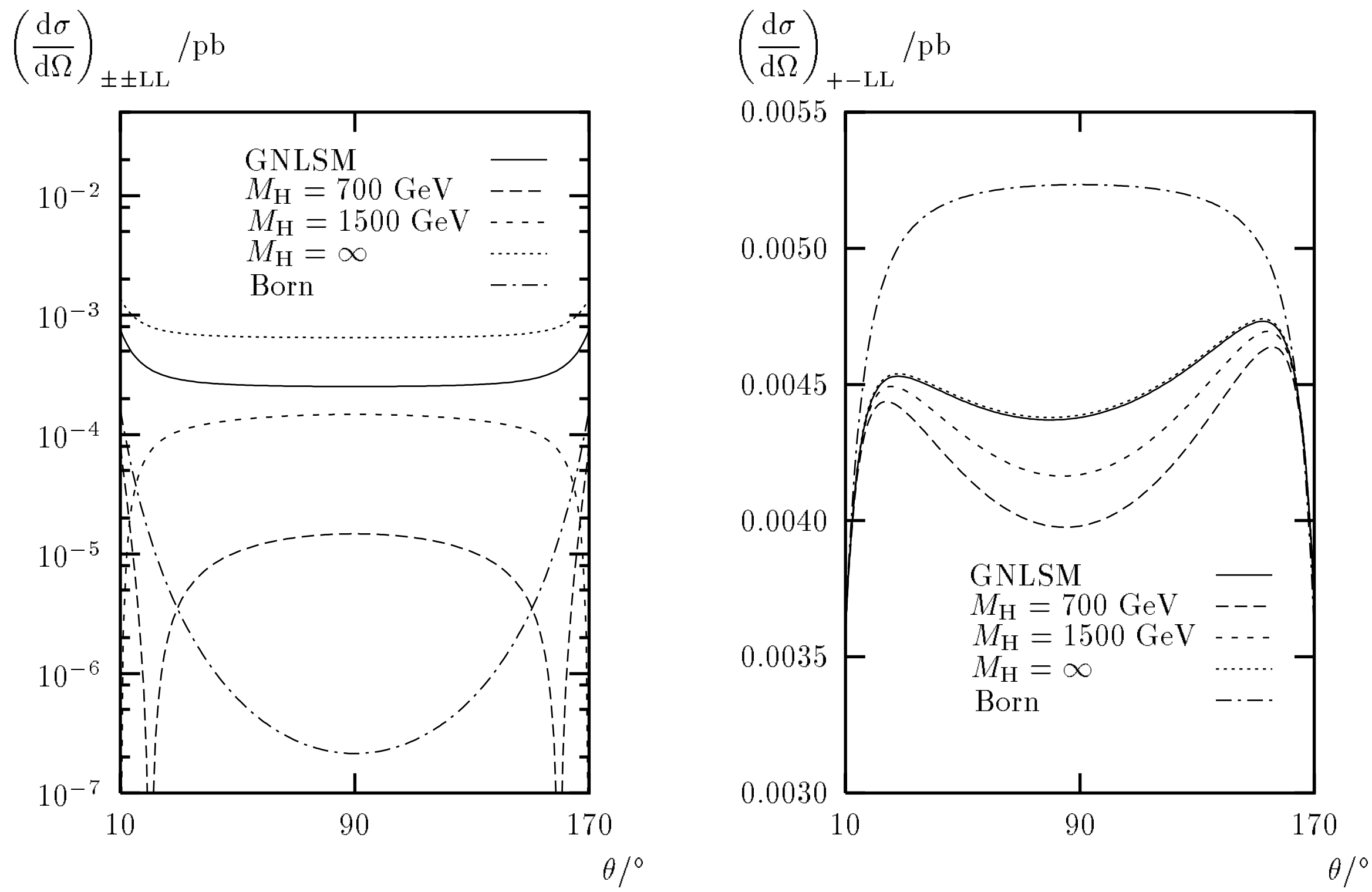

Figure 6: Differential cross-section for purely longitudinal W bosons and equal photon helicities (left) or unequal photon helicities (right).

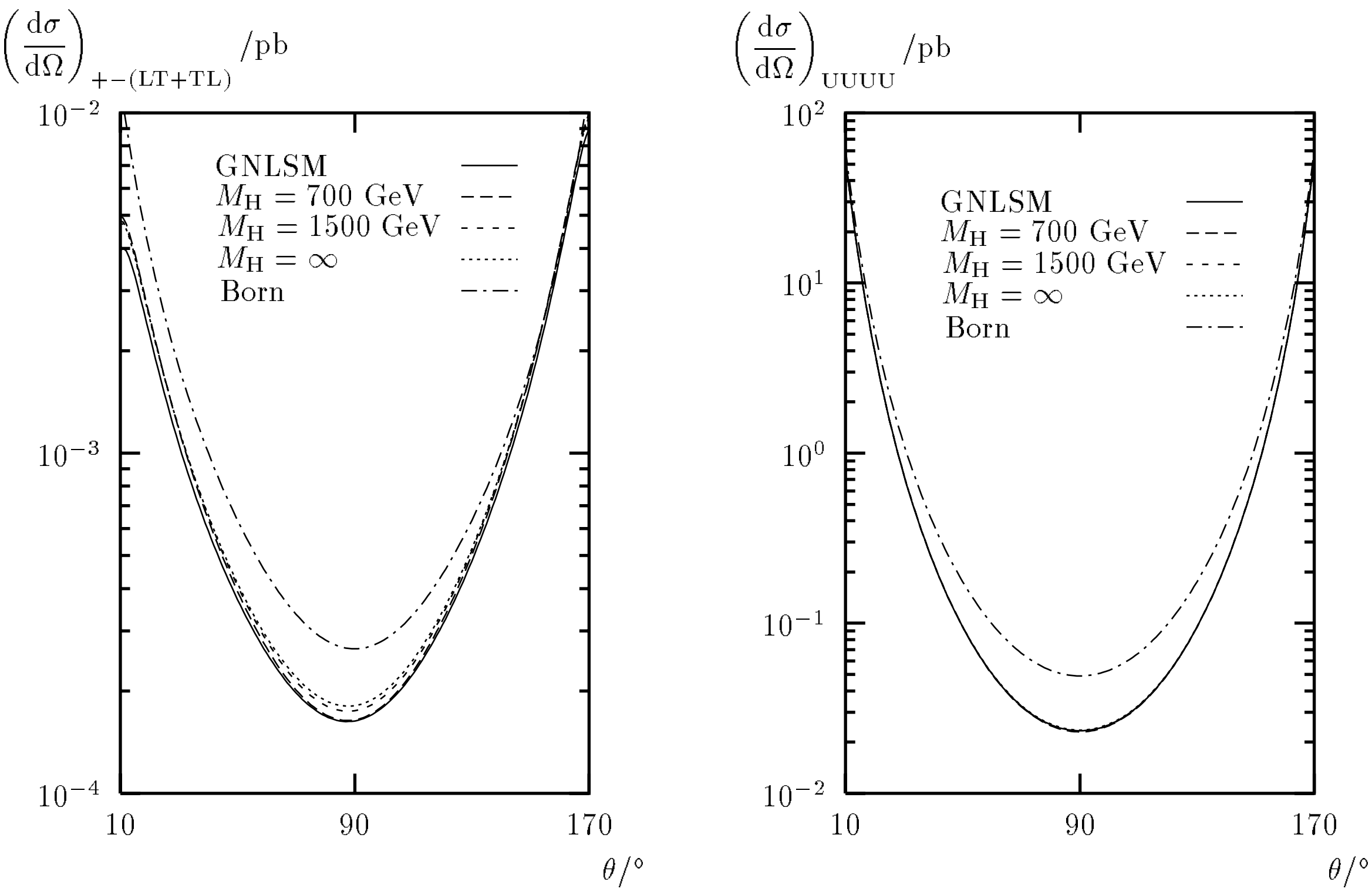

Figure 7: Differential cross-section for for unequal photon helicities and mixed transverse and longitudinal W bosons (left) or unpolarized W bosons and photons (right).

| | $+-$ LL | $+-$ (LT + TL) | $\pm\pm$ UU | UUTT | UUUU |
|---|---|---|---|---|---|
| $\theta = 10°$ | 2% | 38% (55%) | 0.2% | 0.2% | 0.2% |
| $\theta = 90°$ | 13% | 15% (15%) | 2.6% | 0.9% | 3.4% |
| integrated over $10° < \theta < 170°$ | 10% | 2% (9%) | 0.1% | 0.1% | 0.1% |

Table 1: Variation of various polarized cross-sections with the Higgs mass in the range $60\,\mathrm{GeV} < \mathrm{M_H} < \infty$ including the difference to the GNLSM in per cent of the cross-section for $M_\mathrm{H} \to \infty$ at $E_\mathrm{CMS} = 2\,\mathrm{TeV}$.

W bosons is rather flat, the one for transverse W bosons (and thus also for unpolarized ones) is strongly peaked in the forward and backward directions owing to the $t$- and $u$-channel poles at high energies. But even at 90° scattering angle, $\sigma_{\pm\pm\mathrm{LL}}$ is still smaller than $\sigma_{\pm\pm\mathrm{UU}}$ by at least a factor of 50. While $\sigma_{\pm\pm\mathrm{LL}}$ shows a very strong dependence on $M_\mathrm{H}$ and related to that also a sizeable difference between the SM and the GNLSM, the variation of all other polarized cross-sections with $M_\mathrm{H}$, which is in general maximal at 90°, is comparably small. In particular, for unpolarized W bosons it is so small, that in Fig. 7 the SM curves for the various values of $M_\mathrm{H}$ coincide with the one of the GNLSM and that only the lowest-order cross-section can be distinguished. Note furthermore that for $\sigma_{\pm\mp(\mathrm{LT+TL})}$ the curves for $M_\mathrm{H} = 700\,\mathrm{GeV}$ and the GNLSM in Fig. 7, and for $\sigma_{\pm\mp\mathrm{LL}}$ the curves for $M_\mathrm{H} = \infty$ and the GNLSM in Fig. 6 can hardly be separated.

Finally we give in Table 1 some numbers for the variation of the SM corrections with $M_\mathrm{H}$ in per cent of the cross-section for $M_\mathrm{H} \to \infty$. While the deviation between the SM and the GNLSM in general is covered by this range it is given separately for $\sigma_{\pm\mp(\mathrm{LT+TL})}$ in parantheses. The numbers confirm that the cross-sections involving two transverse W bosons depend hardly on the Higgs sector apart from the region close to 90° where the cross-sections are small.

## 5 Conclusions

We have calculated the one-loop radiative corrections to $\gamma\gamma \to \mathrm{W^+W^-}$ in the SM and the GNLSM. Despite of the non-renormalizability of the GNLSM the latter turn out to be UV-finite. The same holds for the limit $M_\mathrm{H} \to \infty$ of the SM corrections since the (logarithmic) one-loop divergences of the GNLSM and the $\log M_\mathrm{H}$ terms in the SM are directly related. However, the complete one-loop results differ by finite terms.

The unitarity-violating effects, which are also different in the SM and the GNLSM, appear only for equal helicities of the incoming photons and purely longitudinal W bosons. The corresponding cross-section depends strongly on the Higgs-boson mass and changes noticeably when going from the SM to the GNLSM. On the other hand, it is strongly suppressed with respect to the one for purely transverse W bosons. The cross-section for transverse W-boson production and also the one for unpolarized W-boson production hardly depend on $M_\mathrm{H}$ and on the realization of the Higgs sector.

In Ref. [8] it was demonstrated that a SM Higgs boson of mass $M_\mathrm{H} \sim 200\,\mathrm{GeV}$ can be seen in $\gamma\gamma \to \mathrm{W^+W^-}$ as a resonance dip in the cross-section. Here we added the results for Higgs masses of several hundred GeV up to the TeV range. Heavy-Higgs effects will only be significant if longitudinally polarized W bosons can be isolated which seems to be extremely difficult owing to the huge background of transversely polarized ones. On the other hand, producing transverse W bosons via $\gamma\gamma \to W^+W^-$ turns out to be practically independent of the mechanism of spontaneous symmetry breaking so that these channels are well-suited for the investigation of other features such as anomalous $\gamma$WW couplings.

**Acknowledgement**

The authors would like to thank C. Grosse-Knetter and H. Spiesberger for useful discussions.